\documentclass[nohyperref]{article}

\usepackage{microtype}
\usepackage{graphicx}
\usepackage{subfigure}
\usepackage{booktabs} 

\usepackage{hyperref}

\usepackage{algorithmic}

\usepackage[accepted]{icml2022}

\usepackage{amsmath}
\usepackage{amssymb}
\usepackage{mathtools}
\usepackage{amsthm}

\usepackage{kotex}
\usepackage{color}
\usepackage{caption}
\usepackage{multirow}

\usepackage[capitalize,noabbrev]{cleveref}

\theoremstyle{plain}

\theoremstyle{definition}

\theoremstyle{remark}

\newcommand{\norm}[1]{\left\lVert#1\right\rVert}
\usepackage[textsize=tiny]{todonotes}

\icmltitlerunning{Guided-TTS: A Diffusion Model for Text-to-Speech via Classifier Guidance}

\begin{document}

\twocolumn[
\icmltitle{Guided-TTS: A Diffusion Model for Text-to-Speech via Classifier Guidance}



\icmlsetsymbol{equal}{*}

\begin{icmlauthorlist}
\icmlauthor{Heeseung Kim}{equal,snu}
\icmlauthor{Sungwon Kim}{equal,snu}
\icmlauthor{Sungroh Yoon}{snu,ai}
\end{icmlauthorlist}

\icmlaffiliation{snu}{Data Science and AI Lab., Seoul National University}
\icmlaffiliation{ai}{Department of ECE and Interdisciplinary Program in AI, Seoul National University}

\icmlcorrespondingauthor{Sungroh Yoon}{sryoon@snu.ac.kr}

\icmlkeywords{Denoising Diffusion, Text-to-speech, Speech, Diffusion, Classifier Guidance}

\vskip 0.3in
]



\printAffiliationsAndNotice{\icmlEqualContribution} 

\begin{abstract}
We propose Guided-TTS, a high-quality text-to-speech (TTS) model that does not require any transcript of target speaker using classifier guidance. Guided-TTS combines an unconditional diffusion probabilistic model with a separately trained phoneme classifier for classifier guidance. Our unconditional diffusion model learns to generate speech without any context from untranscribed speech data. For TTS synthesis, we guide the generative process of the diffusion model with a phoneme classifier trained on a large-scale speech recognition dataset. We present a norm-based scaling method that reduces the pronunciation errors of classifier guidance in Guided-TTS. We show that Guided-TTS achieves a performance comparable to that of the state-of-the-art TTS model, Grad-TTS, without any transcript for LJSpeech. We further demonstrate that Guided-TTS performs well on diverse datasets including a long-form untranscribed dataset.\end{abstract}

\section{Introduction}\label{sec:introduction}

Neural text-to-speech (TTS) models have been achieved to generate high-quality human-like speech from given text \cite{van2016wavenet,shen2018natural}. In general, TTS models are conditional generative models that encode text into a hidden representation and generate speech from the encoded representation. Early TTS models are autoregressive generative models that generate high-quality speech but suffer from a slow synthesis speed due to the sequential sampling procedure \cite{shen2018natural,li2019neural}. Owing to the development of non-autoregressive generative models, recent TTS models can generate high-quality speech with faster inference speed \cite{ren2019fastspeech,ren2021fastspeech,kim2020glow,Grad-TTS}. Recently, high-quality end-to-end TTS models have been proposed that generate raw waveforms from text at once \cite{kim2021conditional,weiss2021wave,chen21p_interspeech}.

Despite the high quality and fast inference speed of speech synthesis, most TTS models can only be trained if the transcribed data of the target speaker are provided. Although long-form untranscribed data, such as audiobooks or podcasts, is available on various websites, it is challenging to use these speech data as training datasets for existing TTS models. To utilize these untranscribed data, long-form untranscribed speech data has to be segmented into sentences, and each segmented speech should then be accurately transcribed. Since the existing TTS models directly model the conditional distribution of speech given text, the direct usage of untranscribed data remains a challenge.

There have also been approaches using untranscribed speech to adapt the pre-trained multi-speaker TTS model for few-shot TTS synthesis \cite{Verification,yan2021adaspeech}. These adaptive TTS models rely heavily on a pre-trained multispeaker TTS model, which is challenging to train and requires high-quality multi-speaker TTS datasets. Also, due to the difficulties of generalization, they underperform in comparison to high-quality single-speaker TTS models such as Glow-TTS and Grad-TTS \cite{kim2020glow,Grad-TTS} trained on a large amount of transcribed data. 

In this work, we propose Guided-TTS, a high-quality TTS model that learns to generate speech with an unconditional DDPM and performs text-to-speech synthesis using classifier guidance. By introducing a phoneme classifier trained on a large-scale speech recognition dataset, Guided-TTS does not use any transcript of the target speaker for TTS. Trained on untranscribed data, our unconditional diffusion probabilistic model learns to generate mel-spectrograms without context. As the untranscribed data does not have to be aligned with the text sequence, we simply use random chunks of untranscribed speech to train our unconditional generative model. This allows us to build training datasets without extra effort in modeling the speech of speakers for which only long-form untranscribed data is available.

To guide the unconditional DDPM for TTS, we train a frame-wise phoneme classifier on a large-scale speech recognition dataset, LibriSpeech, and use the gradient of the classifier during sampling. Although our unconditional generative model is trained without any transcript, Guided-TTS effectively generates mel-spectrograms given the transcript by guiding the generative process of unconditional DDPM using the phoneme classifier. As mispronunciation through guiding error is fatal for the TTS model, we present norm-based guidance that balances the classifier gradient and the unconditional score during sampling. 

We demonstrate that Guided-TTS matches the performance of publicly available high-quality TTS models on LJSpeech without using LJSpeech transcripts. In addition, Guided-TTS generalizes well for diverse untranscribed datasets, and even for a long-form unsegmented dataset (Blizzard 2013). Furthermore, we show that the norm-based guidance significantly reduces pronunciation errors, which allows our proposed model to have a similar level of pronunciation accuracy as the existing conditional TTS models. We encourage readers to listen to samples of Guided-TTS trained on various untranscribed datasets on our demo page.\footnote{Demo : \href{https://bit.ly/3r8vho7}{https://bit.ly/3r8vho7}}
\section{Background}\label{sec:background}

\subsection{Denoising Diffusion Probablistic Models (DDPM) and Its Variant}

DDPM \cite{pmlr-v37-sohl-dickstein15,DDPM}, which is proposed as a type of probabilistic generative model, has recently been applied to various domains, such as images \cite{dhariwal2021diffusion} and audio \cite{WaveGrad,Grad-TTS}. DDPM first defines a forward process that gradually corrupts data $X_0$ into random noise $X_T$ across $T$ timesteps. The model learns the reverse process, which follows the reverse trajectory of the predefined forward process to generate data from random noise.

Recently, approaches have been proposed to formulate the trajectory between data and noise as a continuous stochastic differential equation (SDE) instead of using a discrete-time Markov process \cite{song2021scorebased}. Grad-TTS \cite{Grad-TTS} introduces SDE formulation to TTS, which we have followed and used. According to the formulation of Grad-TTS, the forward process that corrupts data $X_0$ into standard Gaussian noise $X_T$ is as follows:
\begin{equation}
    \label{forward diffusion}
    dX_t = -\frac{1}{2}X_t\beta_tdt+\sqrt{\beta_t}dW_t,
\end{equation}
where $\beta_t$ is a predefined noise schedule, $\beta_t=\beta_0+(\beta_T-\beta_0)t$, and $W_t$ is a Wiener process. 
\citet{Anderson1982-ny} shows that the reverse process, which represents the trajectory from noise $X_T$ to $X_0$, can also be formulated in SDE, which is defined as follows:
\begin{equation}
    \label{reverse diffusion}
    dX_t=(-\frac{1}{2}X_t-\nabla_{X_t}\log{p_t(X_t)})\beta_{t}dt+\sqrt{\beta_t}d\widetilde{W_t},
\end{equation}

where $\widetilde{W_t}$ is a reverse-time Wiener process. Given the score, the gradient of log density with respect to data (\textit{i.e.,} $\nabla_{X_t}\log{p_t(X_t)}$), for $t \in [0, T]$, we can sample data $X_0$ from random noise $X_T$ by solving Eq. (\ref{reverse diffusion}). To generate data, the DDPM learns to estimate the score using the neural network $s_\theta$ parameterized by $\theta$.

To estimate the score, $X_t$ is sampled from the distribution derived from Eq. (\ref{forward diffusion}), given data $X_0$, which is as follows:
\begin{equation}
    \label{property 1}
     X_t|X_0 \sim \mathcal{N}(\rho(X_0, t), \lambda(t)),
\end{equation}
where $\rho(X_0, t) = {\rm e}^{-\frac{1}{2}\int_0^t\beta_{s}ds}X_0$, and $\lambda(t)=I-{\rm e}^{-\int_0^t\beta_{s}ds}$. 
The score can then be derived from Eq. (\ref{property 1}); $\nabla_{X_t}\log{p_t(X_t|X_0)}=-\lambda (t)^{-1}\epsilon_t$, where $\epsilon_t$ is the standard Gaussian noise used to sample $X_t$ given $X_0$ \cite{Grad-TTS}. To train the model $s_\theta(X_t,t)$ for $\forall t \in [0, T]$, the following loss is used:
\begin{equation}
    \label{loss}
    L(\theta) = \mathbb{E}_{t}\mathbb{E}_{X_0}\mathbb{E}_{\epsilon_t}\big[\norm{s_\theta(X_t,t)+\lambda(t)^{-1}\epsilon_t}_2^2\big],
\end{equation}
which is a L2 loss as in previous works \cite{DDPM,song2021scorebased}.

Using model $s_\theta(X_t,t)$, we can generate sample $X_0$ from noise by solving Eq. (\ref{reverse diffusion}). Grad-TTS generates data $X_0$ from $X_T$ by setting $T=1$ and using a fixed discretization strategy \cite{song2021scorebased}:
\begin{equation}
    \label{discretized reverse diffusion}
    X_{t-\frac{1}{N}} = X_t + \frac{\beta_t}{N}(\frac{1}{2}X_t + \nabla_{X_t}\log{p_t(X_t)}) + \sqrt{\frac{\beta_t}{N}}z_t,
\end{equation}
where $N$ is the number of steps required to solve SDE, $t \in \{\frac{1}{N}, \frac{2}{N}, ..., 1\}$ and $z_t$ is standard Gaussian noise.

\subsection{Classifier Guidance}
DDPM can be guided to generate samples with the desired condition without fine-tuning through the introduction of a classifier. \citet{song2021scorebased} use unconditional DDPM to generate class-conditional images by applying a separately trained image classifier. For conditional generation, the classifier $p_t(y|X_t)$ is trained to classify noisy data $X_t$ as condition $y$.

Discretized SDE for conditional generation can be obtained by replacing the unconditional score $\nabla_{X_t}\log{p_t(X_t)}$ in Eq. (\ref{discretized reverse diffusion}) with a conditional score $\nabla_{X_t}\log{p_t(X_t|y)}$.
\begin{align}
    \label{joint}
    &X_{t-\frac{1}{N}} = X_t + \frac{\beta_t}{N}(\frac{1}{2}X_t + \nabla_{X_t}\log{p_t(X_t|y)}) + \sqrt{\frac{\beta_t}{N}}z_t, \\
    \label{separate}
    &\nabla_{X_t}\log{p_t(X_t|y)}=\nabla_{X_t}\log{p_t(X_t)}+\nabla_{X_t}\log{p_t(y|X_t)}.
\end{align}
If the unconditional score and classifier gradient for the target condition are given, the sample $X_0$ with condition $y$ can be generated using Eq. (\ref{joint}). 

\citet{dhariwal2021diffusion} guide not only unconditional DDPM but also conditional DDPM using a classifier. They introduce a gradient scale $s$ when guiding the DDPM, which is multiplied by the classifier gradient ($s\cdot\nabla_{X_t}\log{p_t(y|X_t)}$) to adjust the scale of it. By using $s>1$, they generate higher-fidelity (but less diverse) samples, which contributes to achieving the state-of-the-art performance for class-conditional image generation.
\section{Guided-TTS}\label{sec:model}
In this section, we present Guided-TTS, which aims to build a high-quality text-to-speech model without any transcript of the target speaker. Whereas other TTS models directly learn to generate speech from text, Guided-TTS learns to model unconditional distribution of speech and generates speech from text using classifier guidance. For classifier guidance, we train an unconditional diffusion model on untranscribed speech data and leverage a phoneme classifier trained on a large-scale speech recognition dataset. To the best of our knowledge, Guided-TTS is the first TTS model to generate speech using an unconditional generative model.

\begin{figure*}[h]
    \centering
    \includegraphics[width=0.85\linewidth]{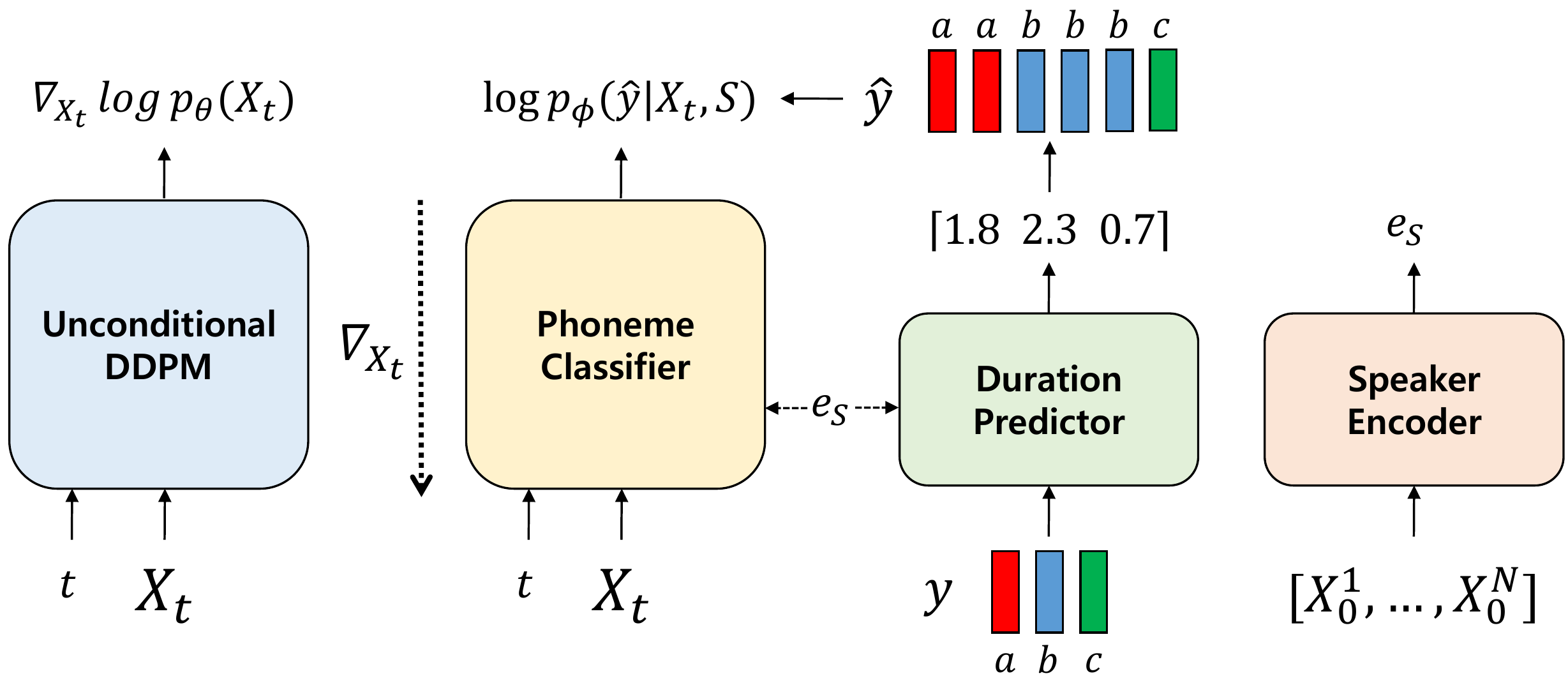}
    \caption{The overall architecture of Guided-TTS. The unconditional DDPM learns to generate speech $X_0$ with untranscribed data. The other modules, the phoneme classifier, duration predictor, and speaker encoder are for guiding the unconditional DDPM to generate conditional samples given $y$.}
    \label{fig1}
\end{figure*}

Guided-TTS consists of four modules: unconditional DDPM, phoneme classifier, duration predictor, and speaker encoder, as shown in Fig. \ref{fig1}. The unconditional DDPM learns to generate mel-spectrogram unconditionally, and the remaining three modules are used for TTS synthesis through guidance. We describe the unconditional DDPM in Section 3.1, followed by the method of guiding the unconditional model for TTS in Section 3.2.

\subsection{Unconditional DDPM}
Our unconditional DDPM models the unconditional distribution of speech $P_X$ without any transcript. We use untranscribed speech data from a single target speaker $S$ as the training data for the diffusion model to build a TTS for the speaker $S$. Since our diffusion model learns without transcript, training samples do not need to be aligned with the transcripts. Thus, we use random chunks of untranscribed speech data as training data such that Guided-TTS does not require speech transcription and sentence-level segmentation when only the long-form untranscribed data is available for the target speaker $S$.

Given a mel-spectrogram $X=X_0$, we define the forward process as in Eq. (\ref{forward diffusion}), which gradually corrupts data into noise, and approximate the reverse process in Eq. (\ref{reverse diffusion}), by estimating the unconditional score $\nabla_{X_t}{\log p(X_t)}$ for each timestep $t$. At each iteration, $X_t, t \in [0, 1]$ is sampled from the mel-spectrogram $X_0$ as in Eq. (\ref{property 1}), and the score is estimated using the neural network $s_\theta(X_t, t)$ parameterized by $\theta$. The training objective of the unconditional model is given by Eq. (\ref{loss}). 

Similar to Grad-TTS \cite{Grad-TTS}, we regard mel-spectrogram as a 2D image with a single channel and use the U-Net architecture \cite{ronneberger2015u} as $s_\theta$. We use the same sized architecture applied to model $32\times32$ sized images in \citet{DDPM} to capture long-term dependencies without any text information, whereas Grad-TTS uses a smaller architecture for the conditional distribution modeling. 

\subsection{Text-to-Speech via Classifier Guidance}
For TTS synthesis, we introduce a frame-wise phoneme classifier and use a classifier guidance method to guide unconditional DDPM. TTS via classifier guidance decouples the generative modeling of speech by conditioning text information. This allows us to leverage a noisy speech recognition dataset as training data for the phoneme classifier, which is challenging to utilize for training existing TTS models. 

As shown in Fig. \ref{fig1}, in order to generate mel-spectrogram given text, our duration predictor outputs the duration for each text token and expands the transcript $y$ to frame-level phoneme label $\hat{y}$. We then sample a random noise $X_T$ of the same length as $\hat{y}$ from the standard normal distribution, and we can generate conditional samples with a conditional score. As in Eq. (\ref{Conditional Score}), we can estimate the conditional score on the left side by adding the two terms on the right side: the first term is obtained from the unconditional DDPM, and the second term can be computed using the phoneme classifier. That is, we build a text-to-speech model with the unconditional generative model for speech by adding the gradient of the phoneme classifier during the generative process.
\begin{equation}
\begin{split}
\label{Conditional Score}
    \nabla_{X_t}{\log{p(X_t|\hat{y}, spk=S)}} &= \nabla_{X_t}{\log{p_{\theta}(X_t| spk=S)}}
    \\&+ \nabla_{X_t}{\log{p_{\phi}(\hat{y}|X_t, spk=S)}}
\end{split}
\end{equation}

To guide the unconditional DDPM for any target speaker $S$, our phoneme classifier and duration predictor are trained on a large-scale speech recognition dataset and designed to be speaker-dependent modules for better generalization to the unseen speaker $S$. We provide the speaker embedding extracted from the pre-trained speaker verification network as a condition for both modules, as described in Fig. \ref{fig1}. We describe each module required for guidance below.

\textbf{Phoneme Classifier}
The phoneme classifier is a network trained on a large-scale speech recognition dataset that recognizes the phoneme corresponding to each frame of the input mel-spectrogram. To train the frame-wise phoneme classifier, we align transcript and speech using a forced alignment tool, the Montreal Forced Aligner (MFA) \cite{MFA}, and extract the frame-level phoneme label $\hat{y}$. The phoneme classifier is trained to classify the corrupted mel-spectrogram $X_t$ sampled from Eq. (\ref{property 1}) as a frame-level phoneme label $\hat{y}$. The training objective of the phoneme classifier is to minimize the expectation of cross-entropy between the phoneme label $\hat{y}$ and output probability with respect to $t\in [0, 1]$.

We use a WaveNet-like architecture \cite{van2016wavenet} as a phoneme classifier, and time embedding $e_t$, which is extracted in the same way as in \citet{Grad-TTS}, is used as a global condition in WaveNet to provide information regarding the noise level of the corrupted input $X_t$ at timestep $t$. For speaker-dependent classification, we also use the speaker embedding $e_S$ from the speaker encoder as the global condition.

\textbf{Duration Predictor}
The duration predictor is a module that predicts the duration of each text token for a given text sequence $y$. We extract the duration label of each text token using MFA for the same data on which the phoneme classifier is trained. The duration predictor is trained to minimize L2 loss between the duration label and the estimated duration in the log-domain, and we round up the estimated duration during inference. The architecture of the duration predictor is the same as that of Glow-TTS \cite{kim2020glow} with the text encoder. We concatenate the text embedding and speaker embedding $e_S$ to predict the speaker-dependent duration.

\textbf{Speaker Encoder}
The speaker encoder encodes the speaker information from the input mel-spectrogram and outputs the speaker embedding $e_S$. Similar to \citet{Verification}, we train a speaker encoder with GE2E loss \cite{GE2E} on the speaker verification dataset and use the speaker encoder to condition speaker-dependent modules. We extract the speaker embedding $e_S$ from the clean mel-spectrogram $X_0$ for each training data. For guidance, we average and normalize the speaker embeddings of the untranscribed speech for the target speaker $S$ to extract $e_S$.

\subsubsection{Norm-based Guidance}
\label{3.2.1}
\begin{algorithm}[h]
  \caption{Norm-based Guidance}
  \label{alg_norm}
\begin{algorithmic}
  \STATE $\hat{y}$: frame-wise phoneme label, $s$: gradient scale, $\tau$: temperature
  \STATE $\theta$: parameter of unconditional DDPM:, $\phi$: parameter of phoneme classifier    
  \STATE $X_1\sim \mathcal{N}(0, \tau^{-1}I)$
  \FOR{$i=N$ {\bfseries to} $1$}
  \STATE $t \leftarrow \frac{i}{N}$
  \STATE $\alpha_t \leftarrow \norm{\nabla_{X_t}\log{p_{\theta}(X_t)}}/\norm{\nabla_{X_t}\log{p_{\phi}(\hat{y}|X_t)}}$
  \STATE $z_t\sim \mathcal{N}(0, \tau^{-1}I)$
  \STATE $\mu_t \leftarrow \frac{1}{2}X_t + \nabla_{X_t}\log{p_{\theta}(X_t)} + s\cdot\alpha_t \nabla_{X_t}\log{p_{\phi}(\hat{y}|X_t)}$
  \STATE $X_{t-\frac{1}{N}} \leftarrow X_t + \frac{\beta_t}{N}\mu_t + \sqrt{\frac{\beta_t}{N}}z_t$
  \ENDFOR 
  \STATE \textbf{return} $X_0$
\end{algorithmic}
\end{algorithm}
Initially, we scaled the gradient of the classifier $\nabla_{X_t}\log{p_{\phi}(\hat{y}|X_t, spk=S)}$ in Eq. (\ref{Conditional Score}) using gradient scale $s$ \cite{dhariwal2021diffusion}.
However, when guiding the unconditional DDPM with the frame-wise phoneme classifier, we found that the norm of the unconditional score suddenly increases near $t=0$ (Appendix \ref{app::norm}). That is, when closer to data $X_0$, the phoneme classifier has little effect on the generative process of the DDPM. As a matter of fact, our experiments on generating samples using various numbers of gradient scale $s$ resulted in mispronouncing samples given text for all cases.

Herein, we propose norm-based guidance to guide the unconditional DDPM better in terms of generating speech conditioned on frame-level phoneme label $\hat{y}$. Norm-based guidance is a method of scaling the norm of the classifier gradient in proportion to the norm of the score in order to prevent the effect of the gradient from being insignificant as the score steeply increases. The ratio between the norm of the scaled gradient and the norm of the score is defined as the gradient scale $s$. By adjusting $s$, we can determine how much the classifier gradient contributes to the guidance of unconditional DDPM. We also use the temperature parameter $\tau$ when guiding the DDPM. We observe that tuning $\tau$ to a value greater than 1 helps generate high-quality mel-spectrograms. Detailed analysis on classifier guidance are in section \ref{5.3}.
\section{Experiments}\label{sec:experiments}
\textbf{Datasets} In Guided-TTS, the speaker-dependent phoneme classifier and duration predictor are trained on LibriSpeech \cite{7178964}, which is a large-scale automatic speech recognition (ASR) dataset with approximately 982 hours of speech uttered by 2,484 speakers with corresponding texts. To extract the speaker embedding $e_S$ from each utterance, we train a speaker encoder on VoxCeleb2 \cite{Voxceleb2}, which is a speaker verification dataset that contains more than 1M utterances of 6112 speakers.

For the comparison case with baselines which make use of the target speaker transcript data, we use LJSpeech \cite{ljspeech17}, a 24-hour female single speaker dataset consisting of 13,100 audio clips. For the other case which makes use of only the untranscribed target speaker speech, we use LJSpeech, Hi-Fi TTS \cite{bakhturina2021hi}, and Blizzard 2013 \cite{King2013TheBC}. Hi-Fi TTS is a multi-speaker TTS dataset with 6 females and 4 males, and the data of each speaker consists of at least 17 hours of speech. We select three relatively clean speakers among them (two males (ID: 6097, 9017) and one female (ID: 92)). Blizzard 2013 is a 147 hours-long audiobook containing both segmented and unsegmented data read by a single female speaker. We use the unsegmented data of Blizzard 2013, randomly clipping 5-seconds-long chunks of audio to build a TTS for long-form untranscribed data.

\textbf{Training Details}
We convert text into International Phonetic Alphabet (IPA) phoneme sequences using open-source software \cite{phonemizer20}. To extract the mel-spectrogram, we use the same hyperparameters as Glow-TTS \cite{kim2020glow}. All modules are trained using Adam optimizer with a learning rate of $0.0001$. For the unconditional model and the phoneme classifier, $\beta_0=0.05$ and $\beta_1=20$ are used for beta schedule. 
Other details and hyperparameters of Guided-TTS are described in Appendix \ref{app::hyperparameters}.

\textbf{Evaluation} To compare the performance of models with transcribed data, we use the official implementations and pre-trained models of Glow-TTS and Grad-TTS.\footnote{Glow-TTS: \href{https://bit.ly/3kS315K}{https://bit.ly/3kS315K}}\footnote{Grad-TTS: \href{https://bit.ly/3qTCmcJ}{https://bit.ly/3qTCmcJ}} For Glow-TTS, we use a pre-trained model with blank tokens between phonemes and use $\tau=1.5$. We use the same hyperparameters as the official implementation, $\tau=1.5$, and the number of reverse steps $N = 50$ for Grad-TTS. To compare model performance in the absence of a transcript, we extract the transcript using a CTC-based conformer-large ASR model \cite{graves2006connectionist,gulati20_interspeech} from NEMO toolkit \cite{kuchaiev2019nemo}, which is pre-trained using LibriSpeech. We train Grad-TTS using the ASR transcribed data for 1.7m iterations, which we refer to as Grad-TTS-ASR. For Guided-TTS, we set $\tau = 1.5$, and the number of reverse steps $N = 50$. We observe that the low classification accuracy of the phoneme classifier near $t=1$ (closer to random noise) deteriorates the sample quality. Therefore, we refrain from using the gradient of the classifier at the initial steps of sampling, setting the gradient scale $s$ to 0. Afterwards, we linearly increase the gradient scale $s$ to $0.3$. For the vocoder, we use the official implementation and pre-trained models of HiFi-GAN.\footnote{HiFi-GAN: \href{https://bit.ly/3FxBv5x}{https://bit.ly/3FxBv5x}}

To show whether Guided-TTS with norm-based guidance generates the sentences of the given text accurately, we measure the character error rate (CER) for each model, which is a metric commonly used in automatic speech recognition (ASR). To compute the metric, we use the CTC-based conformer pre-trained with 7,000 hours of speech from the NEMO toolkit. We generate 5 samples for each sentence in the test set and measure all CER for all generated samples, ultimately using the whole average CER for comparison.
\section{Results}\label{sec:results}
\subsection{Model Comparison}
\label{5.1}
We compare the performances of audio samples by measuring the 5-scale mean opinion score (MOS) on LJSpeech using Amazon Mechanical Turk. In addition, through CER, we check whether the generated sample of each model faithfully reflects the text. To calculate the CER, we first synthesize the speech of a given text for each model and provide it to the ASR model to extract the text corresponding to the generated sample. We then measure the CER between the ground truth text and the text obtained from the ASR model. For evaluation, we randomly select 50 samples drawn from the test set of LJSpeech and measure the MOS and CER.

\begin{table}[t]
\caption{Mean Opinion Score (MOS) with 95$\%$ confidence intervals of TTS models for LJSpeech. "GT MEL" represents the HiFi-GAN result of ground truth mel-spectrogram.}
\label{mos_trans}
\vskip -0.1in
\begin{center}
\begin{small}
\begin{sc}
\begin{tabular}{lccc}
\toprule
\bf Method  &\bf LJ Transcript&\bf 5-scale MOS& \bf CER(\%)\\
\midrule
GT  && 4.45 $\pm$ 0.05&0.64\\
GT Mel && 4.24 $\pm$ 0.07&0.77\\
Glow-TTS   &$\surd$& 4.14 $\pm$ 0.08&0.66\\
Grad-TTS    &$\surd$& 4.25 $\pm$ 0.07&1.09\\
Guided-TTS &$\times$& 4.25 $\pm$ 0.08&1.03\\
\bottomrule
\end{tabular}
\end{sc}
\end{small}
\end{center}
\vskip -0.1in
\end{table}

In Table \ref{mos_trans}, we compare the performance and CER of Guided-TTS with Glow-TTS and Grad-TTS, which are high-quality TTS models. While Glow-TTS and Grad-TTS use transcribed data of LJSpeech, Guided-TTS only uses untranscribed data of LJSpeech to train unconditional DDPM. Guided-TTS shows comparable performance to other TTS models without any transcript of LJSpeech by leveraging the phoneme classifier trained on LibriSpeech. Guided-TTS also has a similar CER to that of the conditional TTS models, which shows that the unconditional DDPM accurately generates speech from the given transcripts using norm-based classifier guidance. This demonstrates that our proposed model enables the building of a high-quality TTS model without any transcript of the target speaker. Samples of all models are available on the demo page.\footnote{Demo : \href{https://bit.ly/3r8vho7}{https://bit.ly/3r8vho7}}

\subsection{Generalization to Diverse Datasets}
\label{5.2}
In the previous section, we showed that Guided-TTS can synthesize high-quality speech without transcript of LJSpeech. Since we separate the training of the unconditional model and the classifier, we are capable of building TTS models for various untranscribed datasets by combining the single phoneme classifier to various unconditional DDPMs trained on untranscribed datasets.

In this section, we assume that only untranscribed speech is available for each speaker. Since existing TTS models inevitably require data with transcripts for training, we extract transcripts from the various untranscribed datasets using a pre-trained ASR model in order to train the powerful baseline, Grad-TTS. We refer to this baseline as Grad-TTS-ASR. Since Guided-TTS leverages the phoneme classifier trained on LibriSpeech, the specific ASR model pre-trained on LibriSpeech is selected to extract transcriptions, making fair comparison possible. For various datasets, we compare Guided-TTS with Grad-TTS-ASR. We use 50 randomly chosen sentences from the test set of each dataset.

The performance of each model on LJSpeech and Hi-Fi TTS is presented in Table \ref{mos_asr}. For LJSpeech, both Guided-TTS and Grad-TTS-ASR achieve comparable performances to Grad-TTS using transcript. However, for Hi-Fi TTS, Guided-TTS outperforms Grad-TTS-ASR and exhibits low CER values for all datasets. This shows that the single phoneme classifier of Guided-TTS stably generates the given text for various datasets. On the other hand, we confirm that the pronunciation accuracy and sample quality of Grad-TTS-ASR, which uses the noisy transcript generated by ASR, are not robust to dataset. Aside from this, we demonstrate that Guided-TTS can robustly generate out-of-distribution (OoD) text for several datasets. The results and details of generated OoD texts are provided in Appendix \ref{app::ood}.
\begin{table}[t]
\caption{Mean Opinion Score (MOS) with 95$\%$ confidence intervals of TTS models for multiple datasets. "Data" refers to the untranscribed speech dataset used for each model. For Blizzard, we use long-form unsegmented data for training.}
\label{mos_asr}
\vskip -0.1in
\begin{center}
\begin{small}
\begin{sc}
\begin{tabular}{clcc}
\toprule
\bf Data&\bf Method&\bf 5-scale MOS&\bf CER(\%)\\
\midrule
&GT&4.45$\pm$0.05&0.64\\
&GT Mel&4.24$\pm$0.07&0.77\\
LJSpeech& Grad-TTS&4.25$\pm$0.07&1.09\\
& Grad-TTS-ASR&4.23$\pm$0.08&1.16\\
&Guided-TTS&4.25$\pm$0.08&1.03\\
\midrule
&GT&4.48$\pm$0.07&0.09\\
Hi-Fi TTS&GT Mel&4.27$\pm$0.07&0.20\\
(ID: 92)& Grad-TTS-ASR&4.11$\pm$0.08&1.33\\
&Guided-TTS&4.20$\pm$0.08&0.81\\
\midrule
&GT&4.50$\pm$0.05& 0.24\\
Hi-Fi TTS&GT Mel&4.26$\pm$0.07&0.33\\
(ID: 6097)&Grad-TTS-ASR&4.09$\pm$0.08&1.88\\
&Guided-TTS&4.16$\pm$0.08&0.79\\
\midrule
& GT&4.45$\pm$0.05&0.11\\
Hi-Fi TTS&GT Mel&4.21$\pm$0.07&0.07\\
(ID: 9017)&Grad-TTS-ASR&3.83$\pm$0.09&2.04\\
&Guided-TTS&4.04$\pm$0.09&0.21\\
\midrule
&GT&4.44$\pm$0.05&0.51\\
Blizzard&GT Mel&4.26$\pm$0.09&0.48\\
&Guided-TTS&4.24$\pm$0.09&0.24\\
\bottomrule
\end{tabular}
\end{sc}
\end{small}
\end{center}
\vskip -0.2in
\end{table}
We also show the performance of Guided-TTS trained with random chunks of unsegmented data of Blizzard 2013, a long-form audiobook dataset, in Table \ref{mos_asr}. Guided-TTS generates high-quality samples without the transcript of Blizzard dataset, just like it has done with the other datasets. In addition, the low CER of Guided-TTS indicates that a TTS model of accurate pronunciation can be built even when using randomly cropped audio without sentence-level segmentation for training.

Based on the results above, we demonstrate that the proposed method enables TTS for untranscribed datasets of various characteristics (\textit{e.g.,} gender, accent, and prosody). Samples on various speakers are available on the demo page.

\subsection{Analysis}
\label{5.3}
\textbf{Norm-based Guidance}
We also compare the proposed norm-based classifier guidance with the classifier guidance used in previous works \cite{song2021scorebased,dhariwal2021diffusion}. A model that conducts a conditional generation task with classifier guidance occasionally generates samples of conditions other than the target condition \cite{song2021scorebased}. Similarly, we observe that Guided-TTS with the classifier guidance method produces mispronounced samples given text. To show the effect of norm-based guidance and adjustment of the gradient scale, we measure the CER of Guided-TTS for LJSpeech according to the gradient scale $s$. We explore the gradient scale $s$ within $[0.5, 1.0, ..., 5.0]$ for classifier guidance \cite{dhariwal2021diffusion}, and $[0.1, 0.2, ..., 1.0]$ for norm-based guidance. 

\begin{figure}[t]
    \centering
    \includegraphics[width=0.85\linewidth]{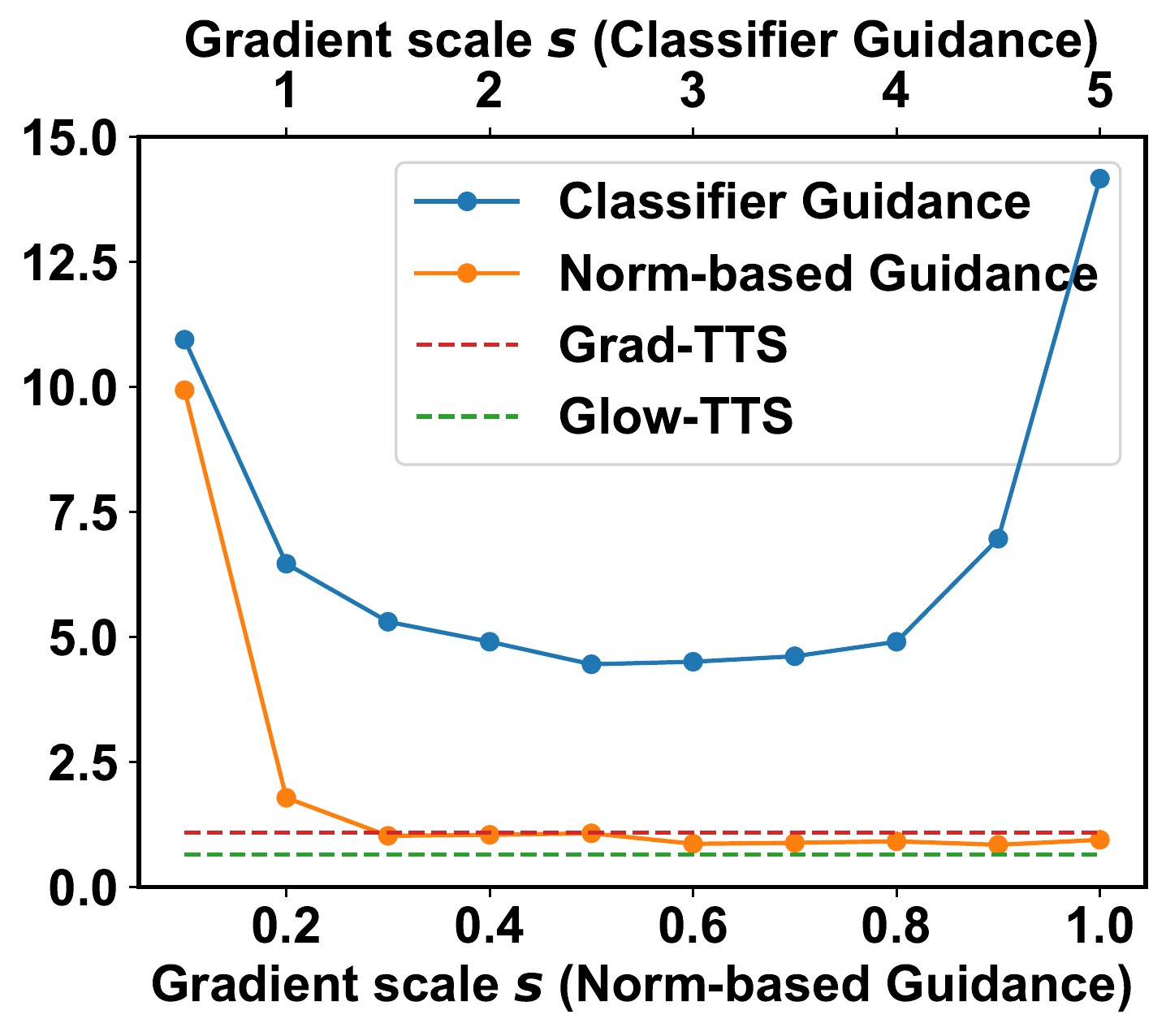}
    \caption{CER of Guided-TTS with classifier guidance \cite{dhariwal2021diffusion} and norm-based guidance according to gradient scales.}
    \label{fig:2}
    \vskip -0.2in
\end{figure}

Fig. \ref{fig:2} presents the CER of Guided-TTS with the classifier guidance \cite{dhariwal2021diffusion} and the proposed norm-based classifier guidance. As shown in Fig. \ref{fig:2}, the sample generated using the existing guidance method shows a far worse CER than the existing TTS models, which indicates that it is unsuitable for TTS. By contrast, the proposed guidance method with the appropriate gradient scale helps accurately generate samples given text sentences, similar to existing TTS models.

If the gradient scale is too small, the effect of the classifier gradient is negligible, and the generated samples do not reflect the given text. On the other hand, we observed that guidance with a large gradient scale deteriorates the sample quality. For the proposed norm-based guidance, we set the default gradient scale $s$ to $0.3$, which generates high-quality samples that exactly match the given text. Samples for multiple gradient scales with each guidance method are on the demo page.

\begin{figure}[h]
    \centering
    \includegraphics[width=0.9\linewidth]{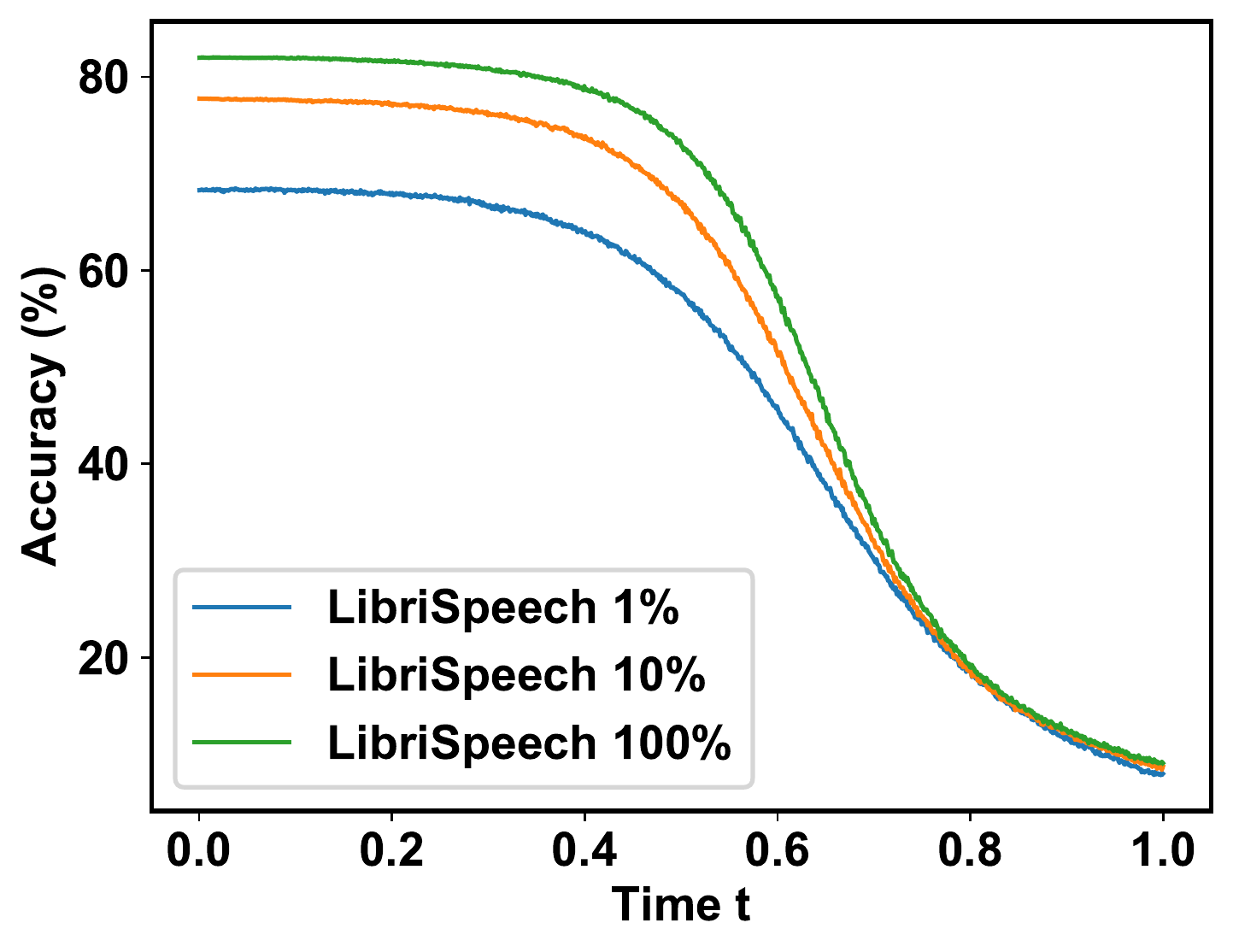}
    \caption{Phoneme classification accuracy on LJSpeech test set over timestep $t$. The number next to the term LibriSpeech indicates the portion of LibriSpeech used for training.}
    \label{fig:acc}
    \vskip -0.15in
\end{figure}

\textbf{Amount of Data for Phoneme Classifier}
We show the CER of Guided-TTS on LJSpeech according to the amount of LibriSpeech data used for training the phoneme classifier in Table \ref{mos_sr_data}. We train the phoneme classifiers with 1\% (9 hours), 10\% (96 hours), 100\% (960 hours) of LibriSpeech respectively, and the classification accuracy of each model is shown in Fig. \ref{fig:acc}. The CER results in Table \ref{mos_sr_data} indicate that the amount of data used for the phoneme classifier is critical regarding the pronunciation accuracy of Guided-TTS. Therefore, the pronunciation of Guided-TTS improves as the amount of data used for phoneme classification increases. Thus, we anticipate that Guided-TTS can be improved even further with a much larger-scale ASR dataset. 
\begin{table}[t]
\caption{CER of Guided-TTS on LJSpeech test set according to the amount of data used for training the phoneme classifier.}
\label{mos_sr_data}
\begin{center}
\begin{small}
\begin{sc}
\begin{tabular}{llc}
\toprule
\bf Method & \bf CER(\%) \\
\midrule
Guided-TTS (LibriSpeech 100\%) & 1.03\\
Guided-TTS (LibriSpeech 10\%) & 2.28\\
Guided-TTS (LibriSpeech 1\%) & 4.24\\
\bottomrule
\end{tabular}
\end{sc}
\end{small}
\end{center}
\vskip -0.15in
\end{table}

\section{Related Work}\label{sec:relatedwork}

\textbf{Unconditional Speech Generation}
In general, the unconditional speech generative model \cite{van2016wavenet, vasquez2019melnet}, which models audio without any information, is more challenging than the conditional generative model that synthesizes speech using text or mel-spectrograms. Several works have attempted to unconditionally generate raw waveforms \cite{van2016wavenet,WaveGAN} or to model the unconditional distribution of latent code or mel-spectrogram of audio \cite{oord2017neural,vasquez2019melnet,10.1162/tacl_a_00430,kharitonov-etal-2022-text} instead of directly modeling raw waveforms. Most existing unconditional models have only been used for unconditional audio modeling and no other purposes. To the best of our knowledge, this is the first application of an unconditional model for TTS with appropriate guidance to enable speech synthesis using untranscribed data from a target speaker.

\textbf{Text-to-Speech Models}
Most text-to-speech (TTS) models are composed of two parts: a model that generates intermediate features (\textit{e.g.,} mel-spectrogram) from text \cite{shen2018natural} and a vocoder, which synthesizes raw waveforms from intermediate features \cite{van2016wavenet}. The autoregressive model is used for the text-to-intermediate feature model \cite{wang17n_interspeech,shen2018natural,ping2017deep,li2019neural} and vocoder \cite{van2016wavenet,kalchbrenner2018efficient} to perform high-quality TTS. To improve the sampling speed of the autoregressive models, flow-based generative models \cite{kingma2018glow} and feed-forward models have been proposed for text-to-mel-spectrogram models \cite{ren2019fastspeech,ren2021fastspeech,kim2020glow,shih2021rad} and vocoders \cite{oord2018parallel,prenger2019waveglow,kimflowavenet}. In addition, variational autoencoder based models \cite{kingma:vae,lee2020bidirectional,liu2021vara}, diffusion based models \cite{DDPM, WaveGrad,DiffWave,Grad-TTS,jeong21_interspeech}, and GAN based models \cite{goodfellow2014generative,Kumar2018melgan,binkowski2019high,kong2020hifi} have been proposed as high-quality speech synthesis models with parallel sampling schemes. End-to-end TTS models have recently been proposed, such as \citet{ren2021fastspeech}, \citet{donahue2021endtoend}, \citet{weiss2021wave}, \citet{kim2021conditional}, and \citet{chen21p_interspeech}. 

Most previous TTS models perform conditional generation tasks using transcribed data of the target speaker. On the other hand, Guided-TTS models unconditional distribution of speech with untranscribed data and generates conditional samples with the pre-trained phoneme classifier. By modeling unconditional distribution of speech, Guided-TTS can utilize long-form untranscribed data of the target speaker without sentence-level segmentation or transcription. 

\textbf{Text-to-Speech with Untranscribed Data} There are two main approaches when building a TTS model without the target speaker’s transcript: fine-tuning based approach and speaker embedding based approach. Both approaches require a pre-trained multi-speaker TTS model. In the fine-tuning based approach \cite{yan2021adaspeech}, the mel-spectrogram encoder is combined with the pre-trained TTS model to fine-tune the model with untranscribed speech of the target speaker. Speaker embedding based approach \cite{NEURIPS2018_4559912e, Verification, casanova21b_interspeech} provides the target speaker's embedding extracted from untranscribed speech to the TTS model for adaptation. These methods require a large-scale multi-speaker TTS dataset, which is difficult to collect and challenging to model the distribution. Also, the performances of these approaches are worse than single speaker TTS models \cite{kim2020glow, ren2021fastspeech, Grad-TTS} trained with the $<$speech, text$>$ pair of the target speaker. On the other hand, instead of using a multi-speaker TTS dataset, we utilize an automatic speech recognition (ASR) dataset to build a TTS model, which is relatively easy to collect. By leveraging the phoneme classifier trained on the ASR dataset, Guided-TTS achieves performance comparable to other TTS models \cite{kim2020glow, ren2021fastspeech, Grad-TTS} with untranscribed data of the target speaker.

There is also an approach that utilizes an untranscribed dataset to extract unsupervised linguistic units and reduces the amount of the paired dataset \cite{ZhangL20-66}. This model focuses on TTS for low-resource languages, while Guided-TTS assumes that a large-scale speech recognition dataset is available and only untranscribed data is given for the target speaker.

\textbf{Diffusion-based Generative Models} DDPM \cite{pmlr-v37-sohl-dickstein15, DDPM} has undergone several theoretical developments \cite{song2021scorebased} and produces high quality samples in many domains \cite{DDPM, dhariwal2021diffusion, WaveGrad, Grad-TTS, Luo_2021_CVPR}. A continuous version of DDPM, an SDE-based model \cite{song2021scorebased, Grad-TTS} is also presented. Thanks to many theoretical and practical breakthroughs \cite{song2021denoising, pmlr-v139-nichol21a}, DDPM has also shown strong performance in speech synthesis \cite{WaveGrad, DiffWave, Grad-TTS, jeong21_interspeech}.

A pre-trained unconditional DDPM can be used for various tasks such as imputation \cite{song2021scorebased}, and controllable generation \cite{song2021scorebased}. In particular, the controllable generation allows \cite{dhariwal2021diffusion} to achieve state-of-the-art performance in class-conditional image generation by guiding the DDPM using a gradient from the classifier trained on the same dataset as DDPM. We introduce the classifier guidance method of unconditional DDPM to text-to-speech synthesis. Our unconditional DDPM and the phoneme classifier can be trained using different datasets, making it possible to build a TTS model with the target speaker's untranscribed speech.
\section{Conclusion}\label{sec:conclusion}
In this work, we present Guided-TTS, a new type of TTS model that generates speech given transcript by guiding the unconditional diffusion-based model for speech. As Guided-TTS models unconditional distribution for speech, we can construct a TTS model using the target speaker's untranscribed data. Thanks to the properties of diffusion-based generative models, our unconditional generative model can generate a speech when a transcript is given by introducing the phoneme classifier trained on LibriSpeech. To the best of our knowledge, Guided-TTS is the first TTS model to leverage the unconditional generative model for speech. We showed that Guided-TTS matches the performance of the previous TTS models on LJSpeech without the transcript. We also showed that Guided-TTS generalizes well to diverse untranscribed datasets with the single phoneme classifier. We believe that Guided-TTS can reduce the burden of constructing training datasets for high-quality TTS.

\section*{Acknowledgements}
We would like to thank Jiheum Yeom and Jaehyeon Kim for their helpful discussions. This work was supported by the National Research Foundation of Korea (NRF) grant funded by the Korea government (MSIT) (No. 2022R1A3B1077720), Institute of Information \& communications Technology Planning \& Evaluation (IITP) grant funded by the Korea government(MSIT) [NO.2021-0-01343, Artificial Intelligence Graduate School Program (Seoul National University)], Institute of Information \& communications Technology Planning \& Evaluation (IITP) grant funded by the Korea government(MSIT) (2022-0-00959), AIRS Company in Hyundai Motor and Kia through HMC/KIA-SNU AI Consortium Fund, and the BK21 FOUR program of the Education and Research Program for Future ICT Pioneers, Seoul National University in 2022.

\bibliography{example_paper}
\bibliographystyle{icml2022}


\clearpage
\appendix
\section{Appendix}

\subsection{Training Details and Hyperparamters} \label{app::hyperparameters}
In this section, we cover the training details and detailed hyperparameters of Guided-TTS. We only use untranscribed data of the various target speakers (LJSpeech, Hi-Fi TTS, and Blizzard 2013) for training unconditional DDPMs and we train the phoneme classifier and duration predictor on LibriSpeech. Alignment labels are required for training the phoneme classifier and the duration predictor, and we train Montreal Forced Aligner (MFA) on LibriSpeech to extract the alignment.

The unconditional DDPMs are trained with batch size 16 for all datasets. The phoneme classifier of Guided-TTS uses a WaveNet-like structure with 256 residual channels and 6 residual blocks stacks of 3 dilated convolution layers, and is trained for 200 epochs with batch size 64. The duration predictor is trained for 20 epochs with batch size 64. The speaker encoder is a two-layer LSTM with 768 channels followed by a linear projection layer to extract 256-dimensional speaker embedding $e_S$, and trained for 300K iterations.

For sampling, we use the last checkpoint for the unconditional DDPM and the speaker encoder. For the phoneme classifier and the duration predictor, we use the checkpoint of the epoch that scores best on its respective metric (validation accuracy for the phoneme classifier and validation loss of the duration predictor).

\subsection{Hardware and Sampling Speed} \label{app::hardware}
We conduct all experiments and evaluations using NVIDIA's RTX A40 with 48GB memory. Although the main objectives of Guided-TTS are not focused on fast inference, it can perform real-time speech synthesis on GPU for $N=50$, which is the number of reverse steps we use for evaluation. We measure the sampling speed of Guided-TTS using a real-time factor (RTF). We also measure how much time it takes to compute the unconditional score ($\nabla_{X_t}\log{p_{\theta}(X_t)}$) and gradient of the classifier ($\nabla_{X_t}\log{p_{\phi}(\hat{y}|X_t)}$). Guided-TTS achieves an RTF of 0.486, of which 0.184 is used to calculate the score and 0.291 is used for classifier gradient calculation.

\subsection{Out-of-Distribution (OoD) Text Robustness} \label{app::ood}
\begin{table}[t]
\caption{Mean Opinion Score (MOS) with 95$\%$ confidence intervals of TTS models for out-of-distribution (OoD) text (LJSpeech test set). "Data" refers to the untranscribed speech dataset used for each model.}
\label{mos_asr_ood}
\begin{center}
\begin{small}
\begin{sc}
\begin{tabular}{clcc}
\toprule
\bf Data&\bf Method&\bf 5-scale MOS&\bf CER(\%)\\
\midrule
Hi-Fi TTS&Grad-TTS-ASR&4.14$\pm$0.08&2.15\\
(ID: 92)& Guided-TTS&4.23$\pm$0.07&0.94\\
\midrule
Hi-Fi TTS&Grad-TTS-ASR&3.99$\pm$0.08&2.49\\
(ID: 6097)&Guided-TTS&4.18$\pm$0.08&0.97\\
\midrule
Hi-Fi TTS&Grad-TTS-ASR&3.91$\pm$0.09&2.74\\
(ID: 9017)&Guided-TTS&4.15$\pm$0.08&0.84\\
\bottomrule
\end{tabular}
\end{sc}
\end{small}
\end{center}
\end{table}

From section \ref{5.2}, we have confirmed that Guided-TTS constructs high-quality TTS models of various speakers. Leveraging the phoneme classifier well-trained on ASR data, Guided-TTS generates high-quality samples with precise pronunciation. Since the phoneme classifier is trained on large-scale ASR data, Guided-TTS generates samples from OoD texts robustly, those of which the model has not seen in the target speaker datasets. In Table \ref{mos_asr_ood}, we show the performance and CER of the OoD samples generated by the Guided-TTS model trained on Hi-Fi TTS. We randomly select 50 sentences from LJSpeech's test set for the OoD text. And for comparison, we use Grad-TTS-ASR trained on Hi-Fi TTS to generate speech corresponding to the OoD text.

Through Table \ref{mos_asr_ood}, we observe that Guided-TTS produces high-quality samples. The CER result of Table \ref{mos_asr_ood} indicates that Guided-TTS generates a sample that faithfully reflects the OoD text. On the other hand, we confirm that Grad-TTS-ASR produces inaccurate samples, showing worse quality for OoD text. Through these results, we demonstrate that Guided-TTS is a TTS model that robustly generates samples for diverse text.

\subsection{Norm of the Unconditional Score and Classifier Gradient} \label{app::norm}
\begin{figure*}[t]
    \centering
    \setlength{\tabcolsep}{10.0pt}
    \begin{tabular}{cc}
        \includegraphics[width=0.40\textwidth]{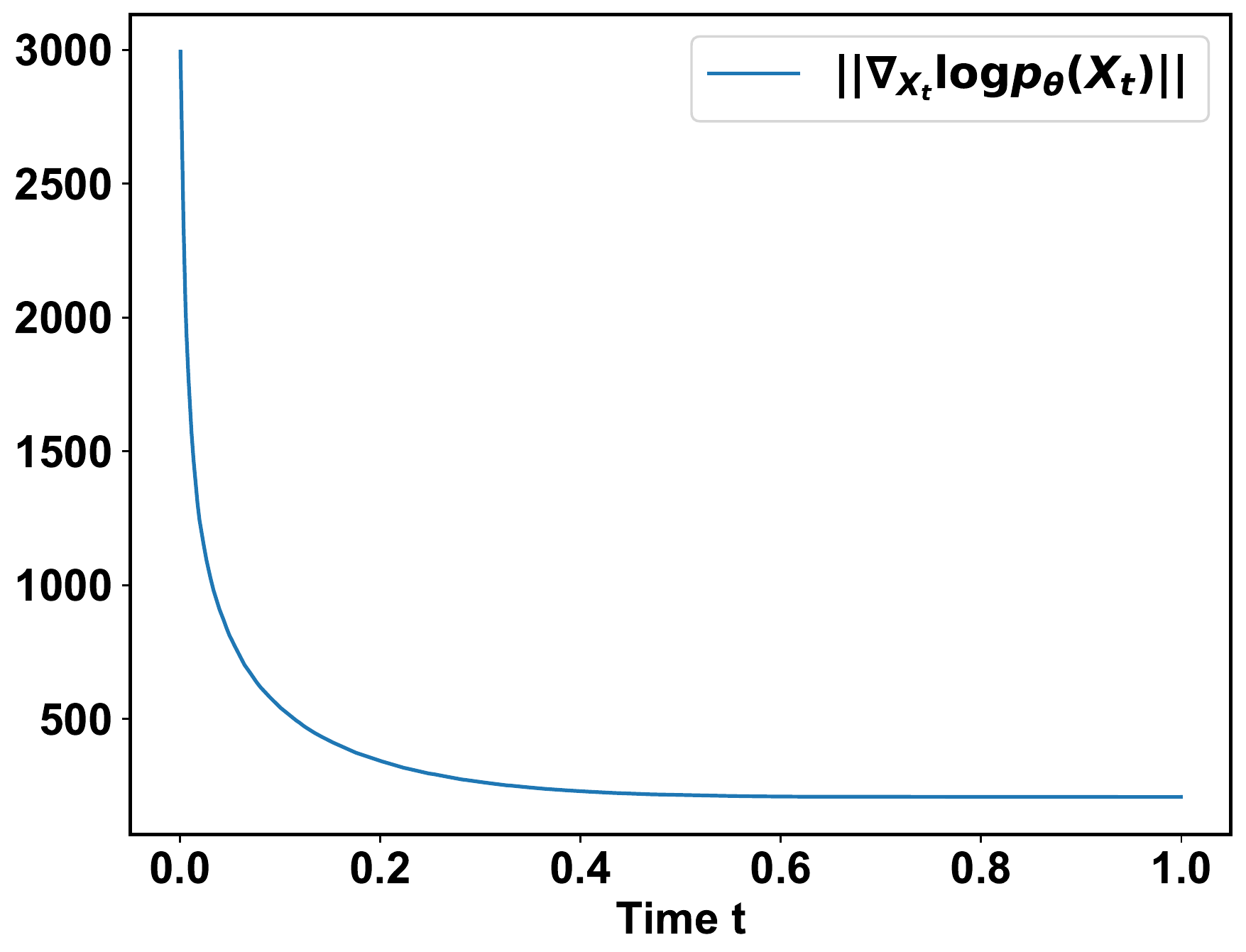} &
        \includegraphics[width=0.40\textwidth]{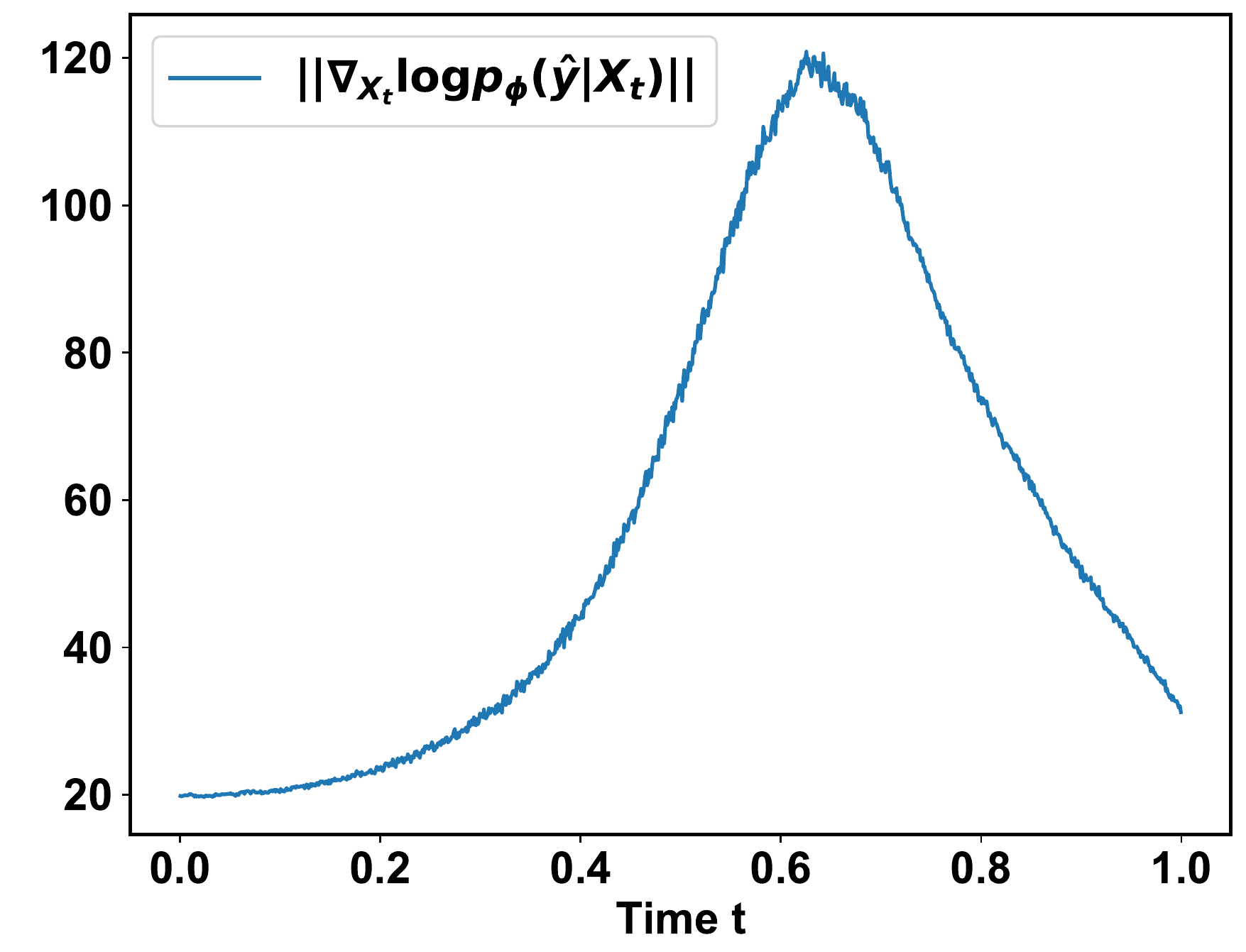} \\
    \end{tabular}
    \caption{The norm of the unconditional score and the classifier gradient for each timestep $t$. (Left) The norm of the unconditional score (Right) The norm of the classifier gradient.}
    \label{fig:norm}
    \vskip -0.1in 
\end{figure*}

The norm of the unconditional score and the gradient norm of the classifier for each timestep are shown in Fig. \ref{fig:norm}. We sample $X_t$ at a total of 1000 timesteps $(t \in (\frac{1}{2000}, \frac{3}{2000}, ..., \frac{1999}{2000}))$ using Eq. (\ref{property 1}) for all 500 samples from the test set of the LJSpeech. We then obtain the norm of the unconditional score and the gradient of the classifier using the sampled $X_t$ with Guided-TTS trained on LJSpeech. Each norm is averaged over 500 samples for each timestep. As shown in Fig. \ref{fig:norm}, the norm of the unconditional score rises steeply around $t=0$. This is about 70 times larger than the norm of the classifier gradient near $t=0$, which significantly reduces the effect of the classifier guidance. To alleviate this problem, we propose the norm-based guidance in Section \ref{3.2.1}, which helps prevent both the gradient of the classifier from being ignored and the issue of synthesized speech not matching the text. 

\subsection{Guided-TTS with Transribed Speech Data} \label{app::guidedt}
Guided-TTS leverages the phoneme classifier trained with LibriSpeech for speech synthesis, taking advantage of training unconditional DDPM and phoneme classifier separately. If transcripts corresponding to the target speaker's speech exist, we can train the phoneme classifier and duration predictor using the target speaker's dataset instead of using LibriSpeech. We refer to this model as Guided-TTS-T. Since Guided-TTS-T uses the same dataset when training the unconditional DDPM and phoneme classifier, there is no need for the phoneme classifier to generalize to unseen speakers, which makes speaker encoder unnecessary for Guided-TTS-T. For comparison, we train all modules in Guided-TTS-T using LJSpeech.

The unconditional DDPM of Guided-TTS-T is trained using the untranscribed speech of LJSpeech in the same way as the unconditional DDPM of Guided-TTS. The phoneme classifier and duration predictor of Guided-TTS-T use the same structure and hyperparameters used in Guided-TTS. We train the phoneme classifier for 1000 epochs and the duration predictor for 60 epochs. Similar to Guided-TTS, we use the checkpoint that scores best on respective metrics (validation accuracy for phoneme classifier, and validation loss for duration predictor) for evaluation.

The performance and CER of Guided-TTS-T are shown in Table \ref{mos_trans_t}. Guided-TTS-T obtains similar performance to Glow-TTS and Grad-TTS even when using the same amount of training data. As demonstrated from this result, Guided-TTS-T is a new approach to construct high-quality TTS in a situation where the target speaker's transcribed data is given.

\begin{table}[t]
\caption{Mean Opinion Score (MOS) with 95$\%$ confidence intervals of TTS models for LJSpeech. "GT MEL" represents the HiFi-GAN result of ground truth mel-spectrogram.} 
\label{mos_trans_t} 
\begin{center}
\begin{small}
\begin{sc}
\begin{tabular}{lccc}
\toprule
\bf Method  &\bf 5-scale MOS& \bf CER(\%)\\
\midrule
GT  & 4.45 $\pm$ 0.05&0.64\\
GT Mel & 4.24 $\pm$ 0.07&0.77\\
Glow-TTS   & 4.14 $\pm$ 0.08&0.66\\
Grad-TTS    & 4.25 $\pm$ 0.07&1.09\\
Guided-TTS-T & 4.23 $\pm$ 0.08&1.21\\
\bottomrule
\end{tabular}
\end{sc}
\end{small}
\end{center}
\end{table}

\subsection{Inpainting} \label{app::inpainting}

\begin{figure*}[h]
    \centering
    \includegraphics[width=0.9\linewidth]{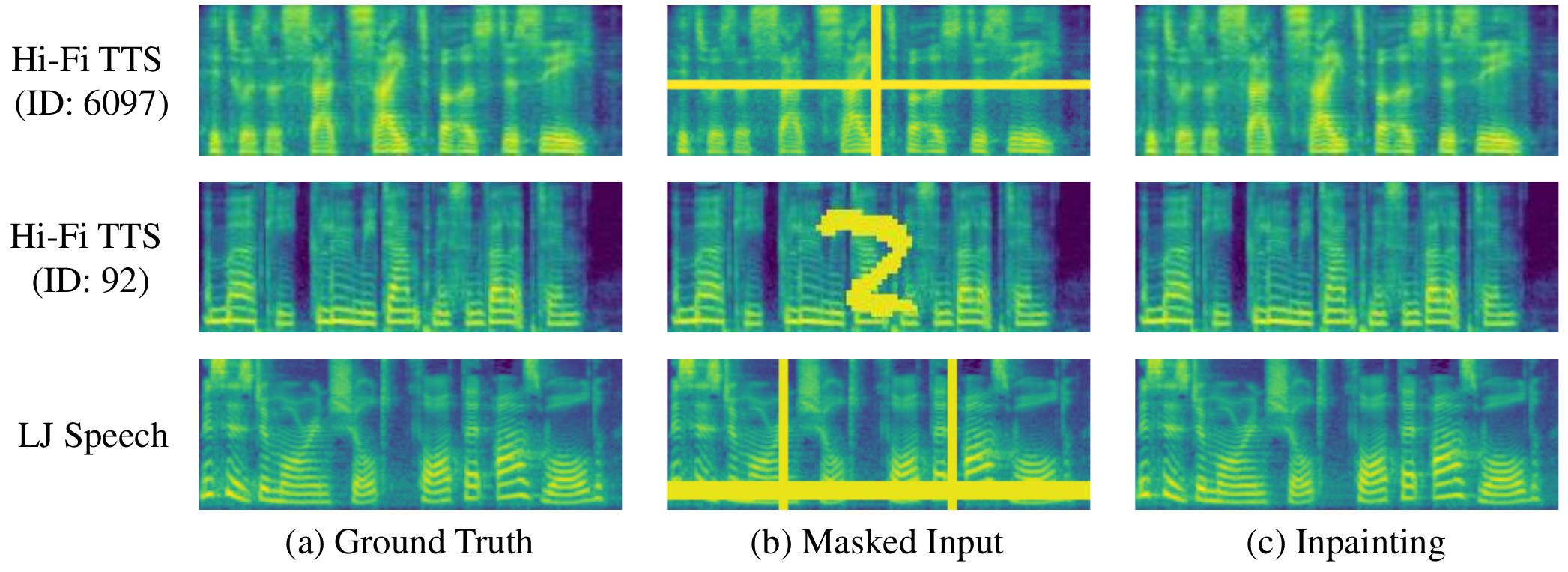}
    \caption{Mel-spectrogram inpainting results of unconditional DDPM trained on LJSpeech, and two speakers (Speaker ID: 92, 6097) from Hi-Fi TTS.}
    \label{fig3}
\end{figure*}

We perform the inpainting task to show how well the unconditional DDPM learns the dependencies in mel-spectrogram. The pre-trained unconditional DDPM fills out the masked part of the mel-spectrogram. We use samples from three speakers; one female speaker (ID: 92), one male speaker (ID: 6097) from Hi-Fi TTS, and a female speaker from LJSpeech. Two cross-shaped masks (LJSpeech, Hi-Fi TTS male) and one binarized MNIST \cite{lecun-mnisthandwrittendigit-2010} mask (Hi-Fi TTS female) are used for masking. We set 1000 as the number of reverse steps $N$ and $\tau=1.5$ for inpainting. The method of inpainting is the same as \citet{song2021scorebased}, and the algorithm is as follows:

\begin{algorithm}[h]
  \caption{Inpainting Mel-spectrogram}
  \label{alg_mas}
\begin{algorithmic}
  \STATE Binary Mask: $M$, Original mel-spectrogram: $\hat{X_0}$ 
  \STATE $\theta$: parameter of unconditional DDPM
  \STATE $X_1\sim \mathcal{N}(0, \tau^{-1}I)$
  \FOR{$i=N$ {\bfseries to} $1$}
  \STATE $t \leftarrow \frac{i}{N}$
  \STATE $\rho(\hat{X_0}, t) \leftarrow {\rm e}^{-\frac{1}{2}\int_0^t\beta_{s}ds}\hat{X_0}$
  \STATE $\lambda(t) \leftarrow I-{\rm e}^{-\int_0^t\beta_{s}ds}$
  \STATE $\hat{X_t}\sim \mathcal{N}(\rho(\hat{X_0}, t), \lambda(t))$
  \STATE $X_t \leftarrow X_t \odot M + \hat{X_t} \odot (1 - M)$ 
  \STATE $z_t\sim \mathcal{N}(0, \tau^{-1}I)$
  \STATE $X_{t-\frac{1}{N}} \leftarrow X_t + \frac{\beta_t}{N}(\frac{1}{2}X_t + \nabla_{X_t}\log{p_\theta(X_t)})+ \sqrt{\frac{\beta_t}{N}}z_t$
  \ENDFOR
  \STATE \textbf{return} $X_0 \odot M + \hat{X_0} \odot (1 - M)$
\end{algorithmic}
\end{algorithm}

The inpainting results are shown in Fig. \ref{fig3}, where (a) is the original mel-spectrogram, (b) is the masked mel-spectrogram, and (c) is the result of inpainting on the masked part. As shown in Fig. \ref{fig3}, we show that the unconditional DDPM of Guided-TTS learns the adjacent frequency and temporal dependencies of the mel-spectrogram. Samples of inpainting results are provided on the demo page.

\end{document}